# GRAVITY+: reducing non-common-path aberrations for sub-10 µas astrometric accuracy

**Quentin Fournier[1*], Guillaume Bourdarot[1], Frank Eisenhauer[1], Helmut Feuchtgruber[1], Dieter Lutz[1], Etienne Rosin[2], Stefan Gillessen[1], Reinhard Genzel[1], Taro Shimizu[1], Oliver Pfuhl[3], Felix Mang[1], Michael Hartl[1], Thomas Ott[1], Ekkehard Wieprecht[1], Vishaal Gopinath[3], Françoise Delplancke-Stroebele[3], Luis Esteras Otal[3], Felix Widmann[1], Sebastiano von Fellenberg[4], Pierre Bourget[3], Guy Perrin[5], Julien Woillez[3] , Paulo Garcia[6], Sebastian Hönig[7], Laura Kreidberg[8], Jean-Baptiste Le Bouquin[9], Thibaut Paumard[5], Christian Straubmeier[10], and GRAVITY+ Collaboration**

[1]Max-Planck-Institut für Extraterrestrische Physik (MPE), Gießenbachstraße, Garching bei München, Germany

[2]CNRS, Sorbonne Universités, France

[3]European Southern Observatory, Karl-Schwarzschild-Str. 2, 85748 Garching, Germany

[4]Canadian Institute for Theoretical Astrophysics, Canada

[5]LESIA, Observatoire de Paris, PSL Research University, CNRS, Sorbonne Universités, UPMC Univ. Paris 06, Univ. Paris Diderot, Sorbonne Paris Cité, 92195 Meudon Cedex, France

[6]Faculdade de Engenharia, Universidade do Porto, rua Dr. Roberto Frias, 4200-465 Porto, Portugal

[7]School of Physics & Astronomy, University of Southampton, Southampton, SO17 1BJ, United Kingdom

[8]Max Planck Institute for Astronomy, Königstuhl 17, 69117 Heidel-berg, Germany

[9]Univ. Grenoble Alpes, CNRS, IPAG, 38000 Grenoble, France

[10]1st Institute of Physics, University of Cologne, Zülpicher Straße 77, 50937 Cologne, Germany

**ABSTRACT**

GRAVITY is a state-of-the-art instrument for near-infrared astrometric interferometry that routinely achieves astrometric accuracy of 30-100 µas in its phase-referenced dual-field mode. However, its fundamental limit is not yet reached, and can still be improved by several factors. In this paper, we focus on the effect of systematics, and in particular the effect of non-common path aberrations between the science channel and the metrology signal in the GRAVITY Fiber Coupler. Through comprehensive laboratory measurements using a dedicated fiber coupler replica and phase-shifting interferometry at 1908 nm, we have characterized these high-order wavefront errors with nanometer precision in the laboratory. We characterized on-sky the effect of these high-order wavefront imperfections on narrow-angle astrometry, using observations of the binary system GJ65. As part of the GRAVITY+ project, high-precision mirrors designed for nanometer-level surface quality, bringing an order of magnitude improvement over existing GRAVITY injection optics, are currently in production and will be installed in the fiber coupler units by the end of 2027. This upgrade is expected to enable astrometric accuracy at the sub-10 μas level in dual-field mode with integration times of just a few minutes.

E-mail: fournierq@mpe.mpg.de

## 1. INTRODUCTION

GRAVITY [1] is a second-generation VLTI instrument operating in the K-band that combines phase-referenced imaging with narrow-angle astrometry capabilities, achieving astrometric precision of tens of microarcseconds and spectral differential astrometry of microarcsecond. Since its commissioning, GRAVITY has made groundbreaking contributions to high-resolution astronomy, in particular in the study of the Galactic Center. These observations enabled the first measurement of gravitational redshift for star S2 orbiting around Sgr A*. The recent detection of S301, a faint main-sequence star with short orbital period and small pericenter distance, offers potential for measuring Sgr A* spin when combining GRAVITY's interferometric measurements with future ELT spectroscopic observations [2]. These observations rely on the unique astrometry accuracy of GRAVITY. In this paper, we aim to investigate the effect of NCPA in dual-field astrometry in order to improve these performances. Here, we investigate the effect of high-order wavefront distortions in GRAVITY Fiber Coupler, characterize in the laboratory, validate their impact on on-sky astrometric data, and propose current avenues to enhance further astrometric accuracy.

## 2. GRAVITY DUAL-FIELD ASTROMETRY

GRAVITY's dual-field observing mode enables narrow-angle astrometry by precisely measuring the angular separation between a science target (SC) and a nearby fringe-tracker object (FT). This capability is implemented through four fiber coupler units [5], one per VLTI telescope. Each fiber coupler employs a roof prism with two facets to split the field of view and direct light through off-axis parabolas (OAPs) into the fibers. The roof prism enables GRAVITY's two distinct observing modes: in OFF-axis operation, two gold-coated facets separate the science target from the fringe-tracking reference, enabling atmospheric piston correction for faint objects; in ON-axis mode, a 50:50 beam splitter and anti-reflective coating allow simultaneous fringe tracking and science observations of the same bright object, optimized for high-resolution spectroscopy. The astrometric measurement relies on GRAVITY's internal metrology system [4], which tracks the optical path difference (OPD) between SC and fringe-tracker FT channels by injecting a 1908 nm laser into both optical paths. The resulting interference pattern at the pupil plane is sampled by four metrology diodes positioned on the telescope's spider arms. The relative phase between FT and SC beams and the metrology signal - for known separation of the telescopes - then gives the angular separation of the two objects on sky.

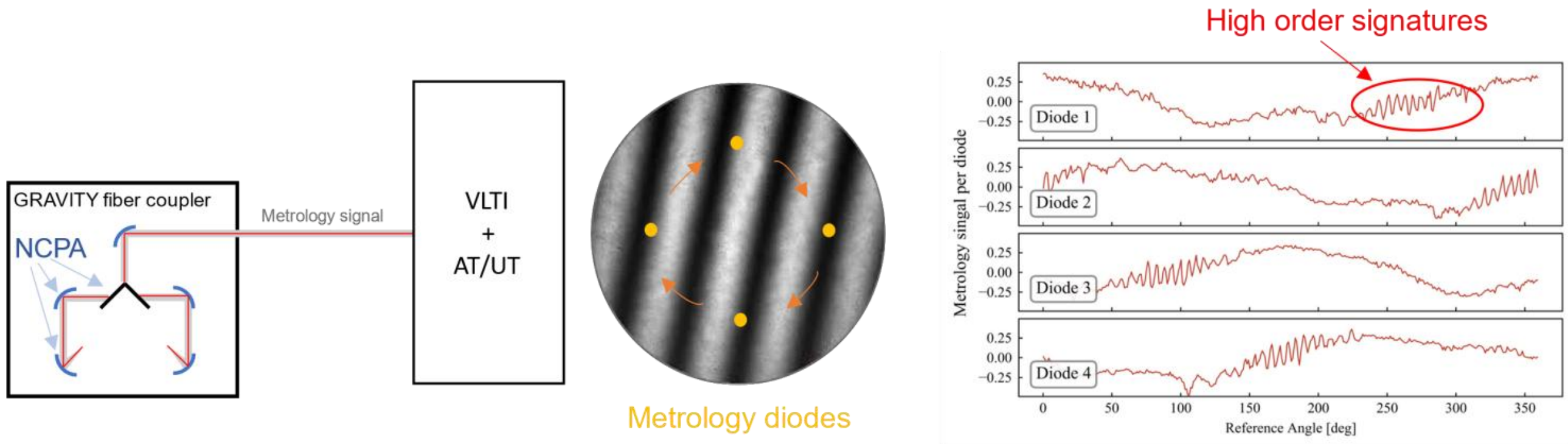

Figure 1: Left: Schematic view of the GRAVITY fiber coupler and metrology. Right: Metrology scan measured on UT2 by the 4 metrological diodes, figure from [6]

However, this methodology approach essentially measures low-order aberrations along the beam path. In addition, the metrology system was further optimized to be insensitive to the defocus term, and largely insensitive to astigmatism, by optimizing the location of the metrology diodes [4]. On the other hand, the sensitivity of the metrology to high-order is constrained by the limited pupil sampling provided by the metrology diodes. Moreover, the system cannot distinguish the non-common path aberrations (NCPAs) that arise specifically in the differential optical path between FT and SC. These NCPAs represent a critical challenge for narrow-angle astrometry, as they directly bias the measured angular separation.

These high-order signatures directly translate into astrometric errors, as shown in Fig 1. These errors were initially attributed to optical defects in the fiber-coupler system, in particular the OAPs. To investigate this hypothesis, we implemented a two-pronged diagnostic approach. First, we characterized the complete wavefront at GRAVITY's output using phase-shifting interferometry (PSI). Second, we isolated the fiber-coupler's contribution by testing a dedicated replica of this subsystem under controlled laboratory conditions. This methodology allows us to disentangle the fiber-coupler's specific impact from other potential sources of aberrations within the VLTI optical train, ultimately identifying the precise optical elements responsible for the observed high-order wavefront errors.

## 3. TWO PHASES OF WAVEFRONT DIAGNOSTICS:

### 3.1. Phase 1: Wavefront measurements at the GRAVITY output

Wavefront sensing measurements were conducted at the GRAVITY instrument's output in the VLTI laboratory using a PSI setup. The metrology laser source, operating at 1908 nm (GRAVITY's metrology wavelength), was injected into both the fringe tracker (FT) and the science channel (SC) fibers. The resulting interference fringes between the FT and SC beams were observed in the pupil plane and imaged onto an infrared (IR) camera.

To temporally encode the fringe pattern, the relative phase of the metrology laser between the FT and SC was modulated at 2 Hz. The wavefront sensor (WFS) data were then demodulated using the Normalization and Orthogonalization Phase-Shifting Algorithm [7] (NOPSA) to reconstruct the optical path difference (OPD) maps across the pupil. High-order aberrations were isolated by removing the first 15 Zernike modes (tip, tilt, defocus, and higher-order terms up to spherical aberration) from the reconstructed wavefront.

The characterization was performed for both on-axis and off-axis observing modes, enabling the assessment of wavefront errors under GRAVITY different operational modes.

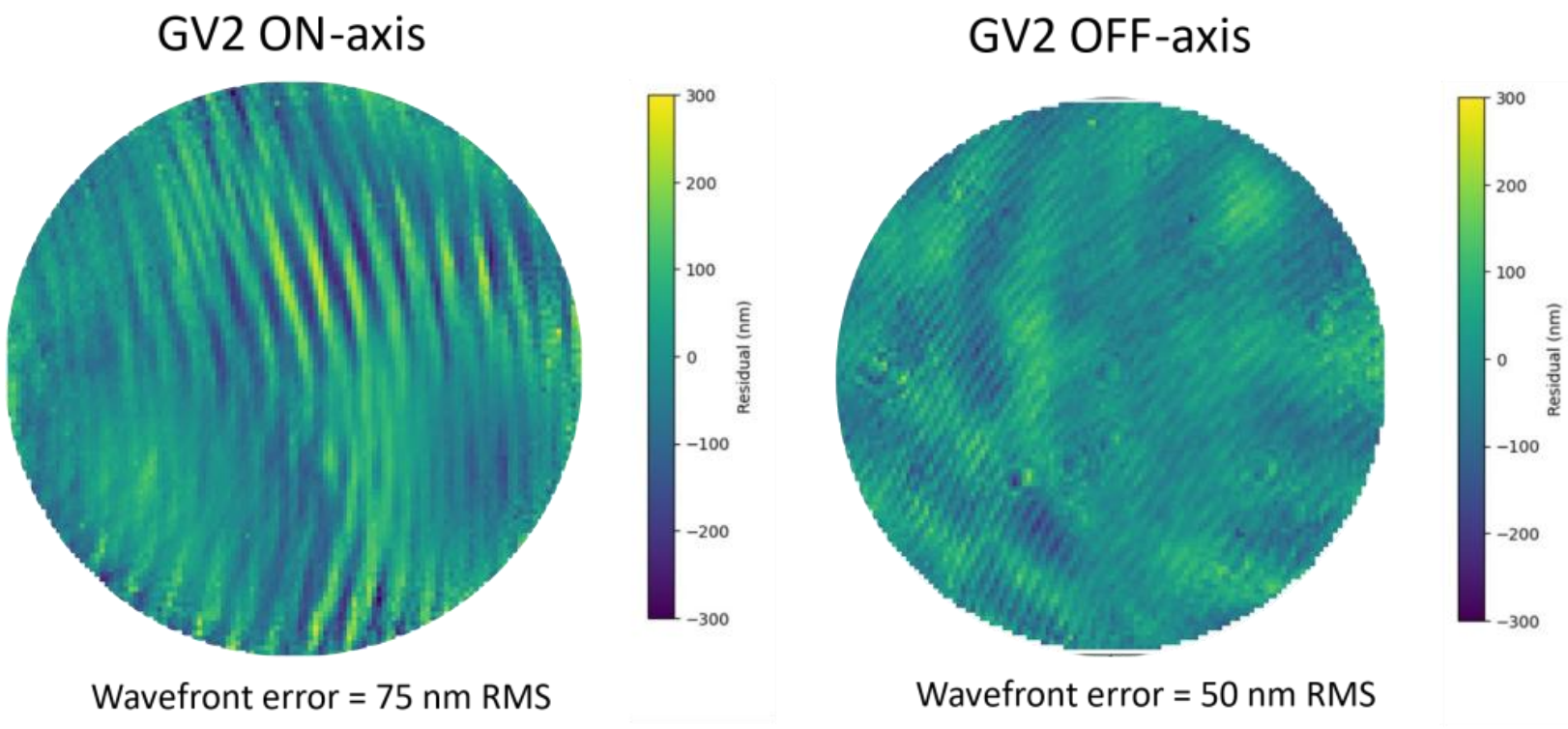

Figure 2: Residual wavefront error maps showing RMS amplitudes of 75 nm (on-axis mode) and 50 nm (off-axis mode).

Residual wavefront errors with RMS amplitudes of 75 nm and 50 nm for on-axis and off-axis respectively were measured in both configurations (example displayed for GV2 in Figure 2). The origin of the signatures seen on the residuals for both modes remain ambiguous based solely on these measurements. However, the observed low-spatial-frequency structures, particularly evident in the off-axis wavefront map, indicate a possible contribution from optical defects in the fiber-coupler system.

### 3.2. Phase 2: Wavefront measurements on a fiber coupler replica

To isolate the contribution of the fiber-coupler optics to the observed wavefront aberrations, a 1:1 replica of the GRAVITY fiber-coupler system was assembled at the Max Planck institute for Extraterrestrial Physics (Garching). The replica incorporates spare optical components from the original units, including mirrors and the roof prism. While functionally identical to the operational GRAVITY couplers, this testbed differs in its actuation system**:** the automated piezo actuators (used for fiber-head positioning, tip-tilt correction, and OPD control) were replaced with manual positioning stages to simplify the experimental setup.

Wavefront measurements were performed using phase-shifting interferometry, following the same principle as the Paranal setup. A fiber-coupled DFB laser ($\lambda$ = 1910 nm) feed both FT and SC arms, one arm passed through an electro-optic phase modulator driven at 2 Hz, while the second arm directly send into the fiber-coupler replica. The fiber-coupler's pupil plane was imaged onto a Xeva MCT 120 infrared camera using a custom-designed lens. The camera operated at a 60 Hz frame rate with frequency reference triggering. The acquired interferograms were processed using the same demodulation routine as in the Paranal measurements (NOPSA, 15 first Zernike are fitted and removed), ensuring consistency in data analysis.

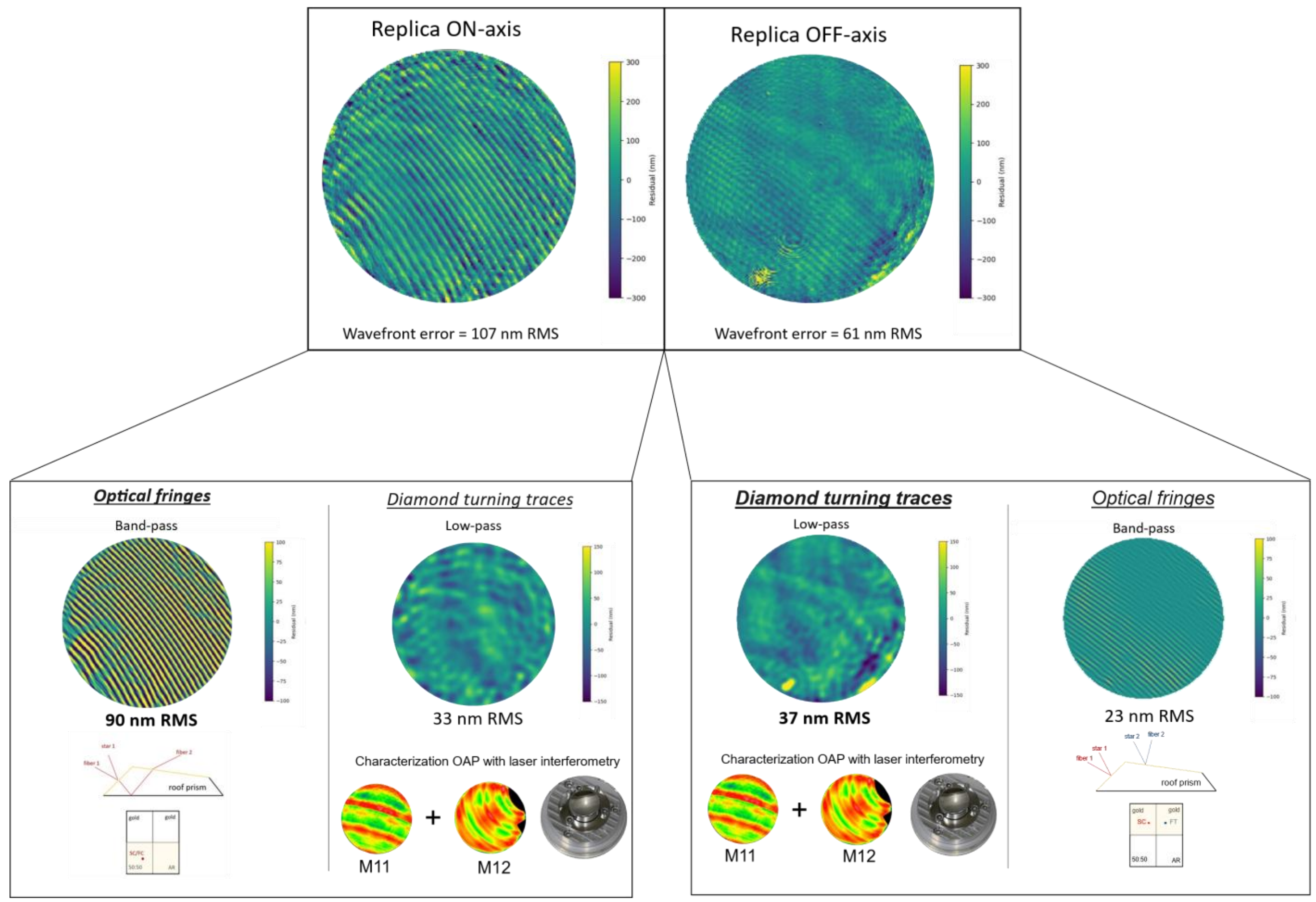

Figure 3: Wavefront residual maps from PSI experiments with the fiber coupler replica, after removal of the first 15 Zernike modes. On-axis residuals show 107 nm RMS wavefront error, primarily (90 nm RMS) from optical fringes in the 50:50 facet coating, with secondary (33 nm RMS) contributions from OAP diamond-turning signatures (combined M11/M12 effects, individually characterized via laser interferometry). Off-axis residuals exhibit 61 nm RMS wavefront error, dominated by OAP diamond-turning signatures (37 nm RMS), with secondary (23 nm RMS) optical fringes from the gold facet coating.

The residual wavefront analysis presented in Figure 3 demonstrates consistent patterns between Phase 1 (Figure 2) and Phase 2 (Figure 3.) measurements, with similar patterns observed in off-axis mode in both measurements. Our investigation reveals two distinct families of aberrations that contribute differentially to the observed wavefront errors depending on the observing mode:

- High-order mirror signatures: The OAPs exhibit surface defects from their diamond-turning manufacturing and polishing process. These defects manifest as millimeter-scale periodic pattern in the spatially filtered phase maps (Figure 3), acting as the primary systematic error source in off-axis mode (37 nm RMS) while remaining a second order contributor in on-axis mode (33 nm RMS). In both cases, the observed signatures closely match those measured on spare M11 and M12 mirrors during individual laser interferometric characterization, confirming their origin. These characteristic patterns do not average for short-exposure time, motivating an upgrade for the post-polishing process of the OAPs.

- Fringing-effect: Optical fringing effects emerge as the dominant contributor in on-axis observations, accounting for 90 nm RMS of the measured residuals, while being a secondary effect in off-axis mode with 23 nm RMS contribution. These fringes appear to originate from resonant cavity effects within the roof prism. This effect is also seen by measuring individual beam intensity map on each channel (FT/SC), showing intensity modulation. In on-axis configuration, we observe a significant contrast difference (in intensity maps) indicative of a multiple-wave interference, for each facets (50/50 coating and AR coating). This contrast approaches unity for the 50:50 beam splitter facet, and approximately 0.1 for the anti-reflective facet. Notably, these high spatial frequency fringes in on-axis mode correlate directly with the characteristic oscillations observed in the metrology diode measurements shown in Figure 1. In off-axis, the contrast measured in intensity maps is similar for both facets (approximately 0.12), indicating a similar origin for both FT/SC channel.

## 4. ON-SKY ANALYSIS

In narrow-angle astrometry, the phase measured can be described as [3]:

$$\frac{2\pi}{\lambda}(\vec{s}-\vec{p})\cdot\overrightarrow{B_{NAB}} = (\phi_{SC}-\phi_{FT}) - \frac{2\pi}{\lambda}D(\lambda_m,\lambda_{SC}) - \text{PHASEMET} + \text{error terms} \quad (1)$$

with $\vec{s}$ and $\vec{p}$ the vector coordinates of the two components of the binary, $\overrightarrow{B_{NAB}}$ the narrow-angle baseline [8], $D(\lambda_m,\lambda_{SC})$ the correction of dispersion between the SC and the metrology, PHASEMET is the phase measured by the metrology and $\lambda$ the wavelength. By definition, NCPA correspond to the differential phase residuals between the stellar light path and the metrology path through the VLTI and GRAVITY systems. The metrology measures the OPD difference between the FT and the SC channels, which directly encodes the on-sky separation between these channels in the absence of OPD perturbations.

When observing OPD residuals, defined as the difference between metrology phase and science phase for an object with known separation, we obtain a direct measurement of NCPA on sky. We performed this test measurement using GJ65, a well-characterized M-dwarf binary system whose separation had been precisely determined with GRAVITY (predicted separation on Figure 6). The observations were conducted during a single night using all four UTs in GRAVITY's high spectral resolution mode, covering both off-axis and on-axis observing modes including a zenith passage.

The observations followed this chronological sequence:

1. An initial off-axis dataset (off-axis 1) comprising 4 exposures with 30° pupil reference angle rotation, including FT/SC swaps
2. An on-axis dataset with 8 exposures covering nearly 60° of pupil rotation
3. A final off-axis dataset (off-axis 2) with 8 exposures covering <10° of pupil rotation

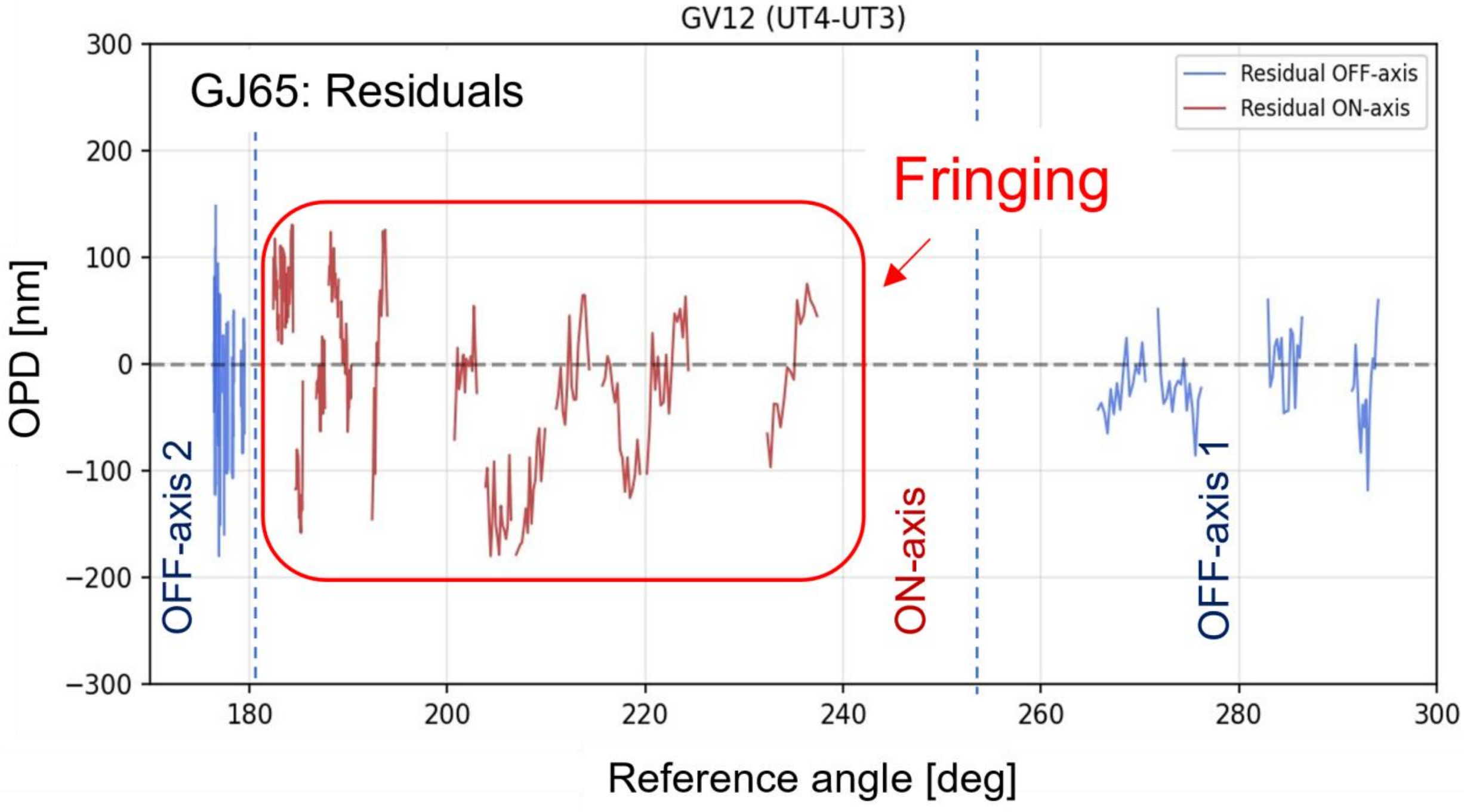

Figure 4: Residual differences between metrology signal and science phase measurements for the three datasets (OFF-axis 1, ON-axis, OFF-axis 2) the baseline GV12. In the ON-axis dataset, high-order aberrations manifest as persistent modulation in the residuals, directly tracing the NCPA contribution.

Figure 4 presents the residual phase differences, which combine our separation estimation errors, NCPA contributions, and measurement noise. Notably, the high SNR in on-axis mode reveals an oscillatory pattern at relatively high frequency (period ~5-10°, amplitude ~300 nm) that emerges clearly above the noise. This periodicity and amplitude match the signatures observed in the metrology diodes (Figure 5), indicating that the dominant fringing effects in on-axis mode significantly impact the OPD residuals.

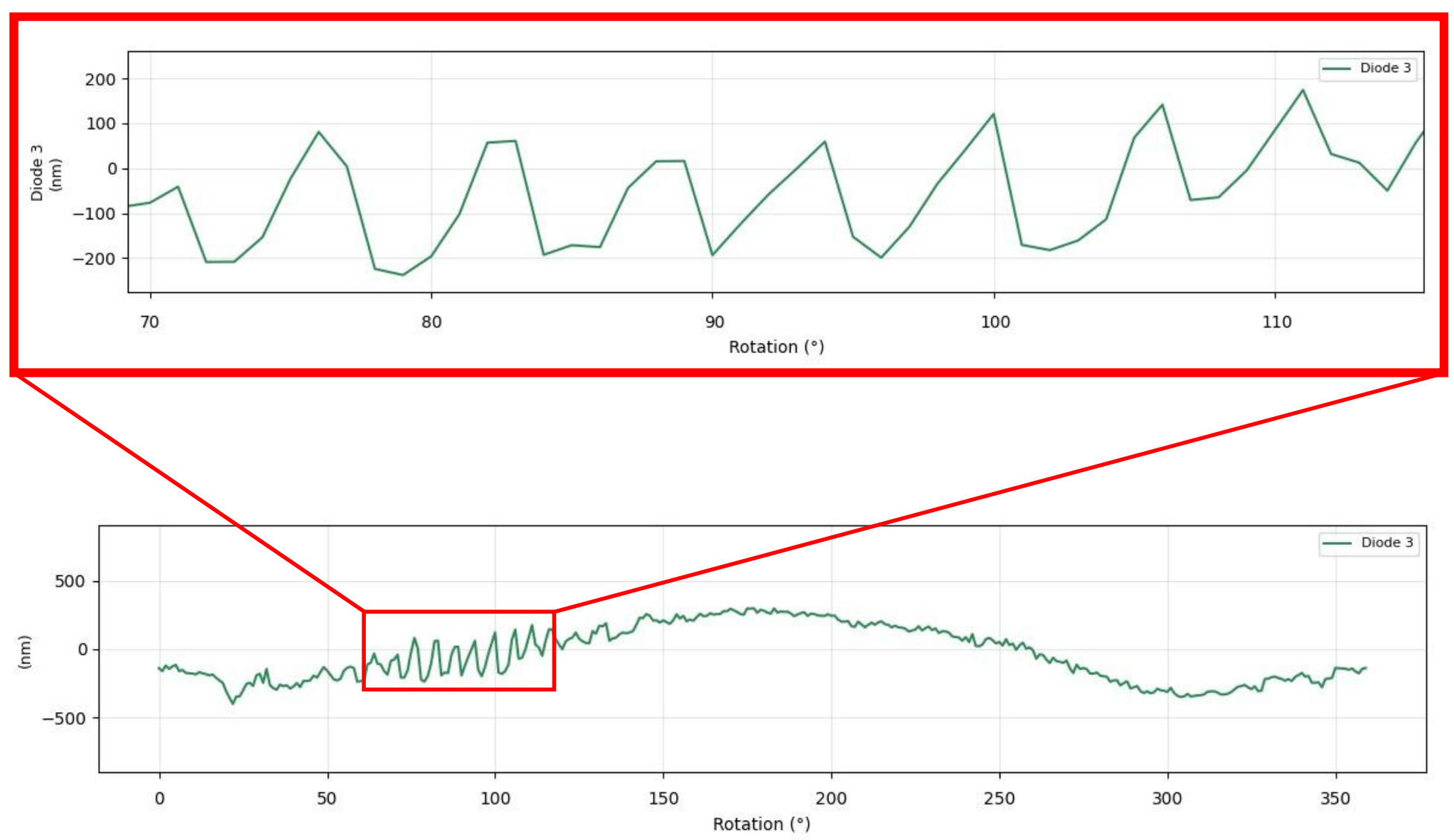

Figure 5: Detail of metrology diode signal in UT on-axis mode showing characteristic wiggles. These oscillations, most likely induced by fringe effects, exhibit a spatial frequency of approximately 7° per cycle with an amplitude of ~300 nm.

These results indicate that, in the absence of sufficient pupil modulation, such systematic effects are difficult to average out over short acquisitions of only a few minutes. This is particularly relevant for exo-GRAVITY observations, for which the on-axis mode is commonly used and short integration sequences are often required [9]. In the first off-axis sequence, despite the relatively large pupil rotation covered during the observations, no significant oscillatory pattern was detected, in contrast to the on-axis data. This may be explained by the smaller amplitude of the fringing effect in off-axis mode, whose RMS is about four times lower than in the on-axis configuration. In this case, the effect may therefore remain below the noise level of the metrology residuals. The second off-axis sequence does not show sufficient pupil rotation, which does not allow us to probe the systematic effects.

The relative astrometry was derived from the residual OPD between the internal metrology signal and the science interferometric phase. After applying a common phase reference to both the science visibilities and the metrology phases, residuals were computed for each of the six VLTI baselines and converted into optical path differences. For the off-axis data, the metrology zero point was calibrated using the swap sequence, in which the two binary components are observed sequentially in the science and fringe-tracking channels. Since the astrometric contribution changes sign between the two configurations whereas the metrology zero point remains common, the average residual of the two swapped states provides a baseline-dependent estimate of the instrumental zero point. The corrected residuals were then expressed in a common astrometric reference frame and interpreted as the delay induced by a small offset between the true binary position and the nominal fiber separation. For each DIT, the six baseline residuals were fitted with a linear astrometric model in which the optical path difference on each baseline is given by the projection of the two-dimensional sky offset onto the corresponding projected baseline. The two fitted parameters are the offsets in right ascension (RA) and declination (Dec). Because six baselines are available for only two astrometric parameters, the system is over-constrained and was solved by least-squares inversion. For the on-axis data, the same fitting procedure was applied after defining the instrumental zero point from the zero on-axis observations. The resulting astrometric correction was determined for each DIT and then averaged over DITs and exposures. The final astrometric separation is therefore deduce adding the imposed fiber separation.

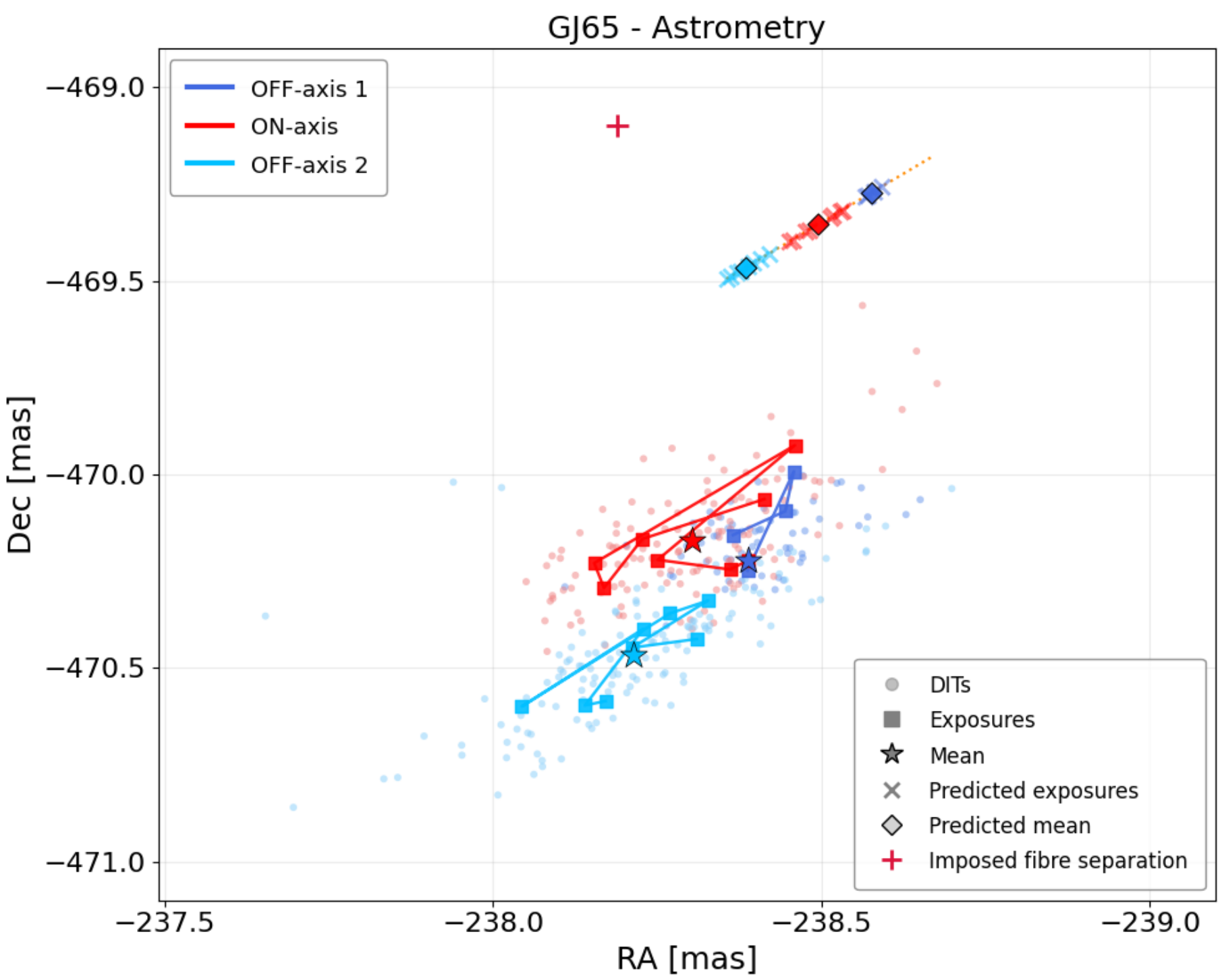

Figure 6: Astrometric separation measurements in right ascension (RA) and declination (Dec) for the three observation datasets: off-axis 1 (dark blue), on-axis (red), and off-axis 2 (light blue). The predicted separation (crosses) and average predicted separation (diamonds) are derived from previous GRAVITY orbital measurements. Individual data points represent astrometric measurements for each DIT, squares show means per exposure, and stars indicate dataset averages. The red cross denotes the imposed fiber separation.

Figure 6 displays the derived astrometric separations across the three measurement phases. While the mean measured values generally follow the predicted trend, a systematic examination reveals a consistent offset between experimental and predicted separations, amounting to approximately 200 µas in right ascension and -1 mas in declination.

| Dataset | $N_{exp}$ | $S_{pred}$ [RA, Dec] (mas) | $S_{exp}$ [RA, Dec] (mas) | $\Delta S$ [RA, Dec] (mas) |
|---|---|---|---|---|
| **Off-axis 1** | 4 | -238.58(1) ; -469.27(1) | -238.41(5) ; -470.12(11) | 0.16(5) ; -0.85(11) |
| **On-axis** | 8 | -238.50(3) ; -469.35(3) | -238.30(12) ; -470.17(12) | 0.19(12) ; -0.82(12) |
| **Off-axis 2** | 8 | -238.38(2) ; -469.47(2) | -238.21(9) ; -470.47(11) | 0.17(10) ; -1.00(11) |

Table 1 : Summary of astrometric separation results showing averaged measurements in right ascension (RA) and declination (Dec) for three observation datasets: off-axis 1 (dark blue), on-axis (red), and off-axis 2 (light blue). The table includes: number of exposures ($N_{exp}$), predicted separation ($S_{pred}$), measured separation ($S_{exp}$), and the difference ($\Delta S$) between predicted and measured values all in mas.

The statistical analysis each dataset in terms of individual exposures demonstrates distinct performance characteristics between observing modes. Declination measurements maintain consistent dispersion around 100 µas across all datasets, whereas right ascension measurements show more pronounced variations. First off-axis observations yield the highest precision, with standard deviations of 50 µas in RA and 110 µas in DEC. In contrast, on-axis measurements exhibit approximately twice the dispersion in RA (120 µas) compared to the initial off-axis dataset, while the second off-axis sequence shows intermediate precision at 90 µas in RA.

The observed discrepancy between predicted and measured dispersion patterns, combined with the separation variations between acquisitions, indicates that statistical dispersion primarily originates from systematic errors rather than actual astrometric changes. This effect is particularly pronounced in on-axis observations, where the previously identified NCPA-induced optical fringes in the metrology system directly impact astrometric measurements. For off-axis observations, a combination of diamond-turning artifacts and residual optical fringing could cause limit measurement accuracy.

These findings underscore the necessity of applying the pupil modulation technique, introduced in 2022, for both GRAVITY observing modes (on- and off-axis). This implementation is already in place for off-axis observations in the Galactic Center, for which it is routinely used in operation. Its use should be investigated for on-axis observations as well, for which the effect of averaging high-order aberrations is even more critical.

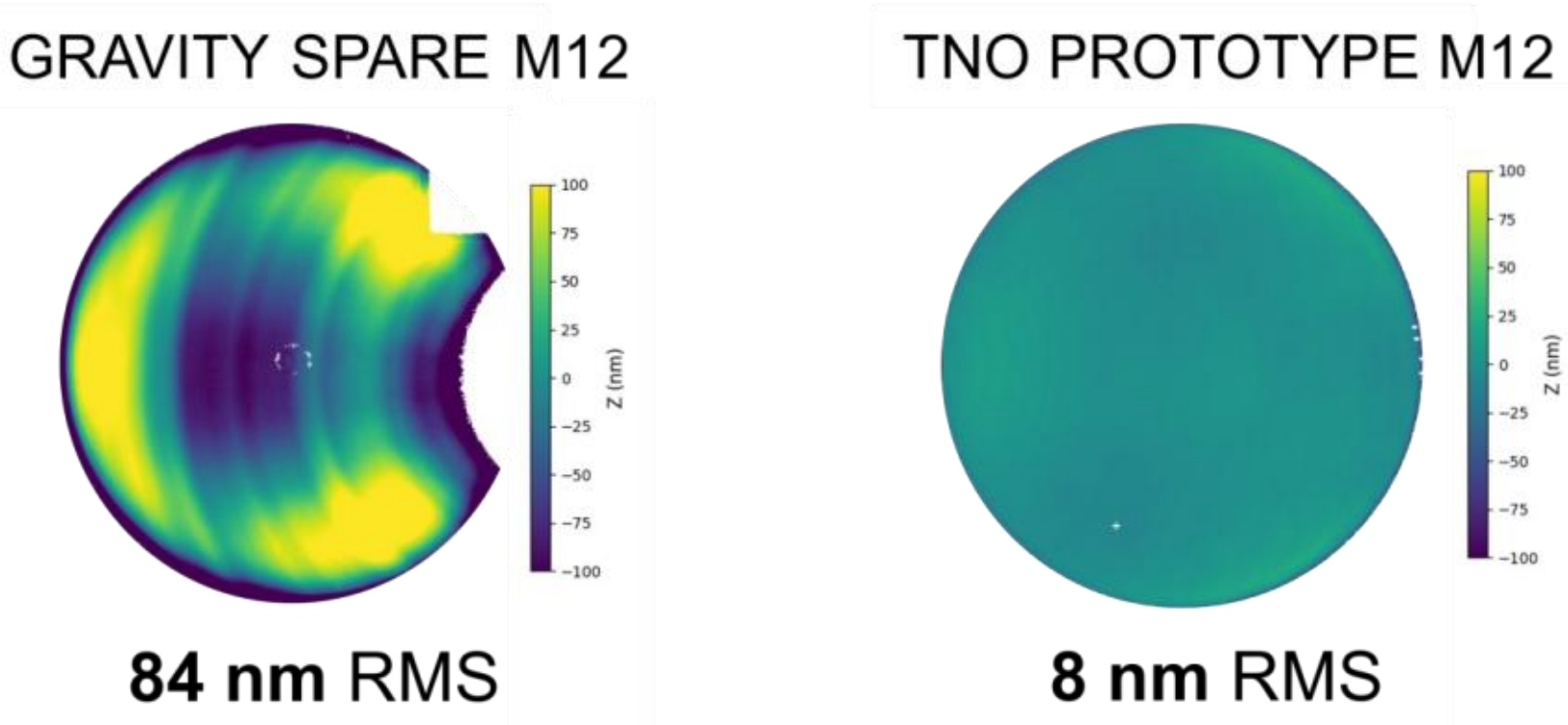

Figure 7: Wavefront residual maps from laser interferometry measurements of GRAVITY fiber coupler components (M12) after removal of the first 6 Zernike modes (same colorscale). The current GRAVITY M12 (left) exhibits diamond-turning marks and trefoil signatures with 84 nm RMS surface error, while the TNO-provided prototype (right) shows improved surface quality with 8 nm RMS error.

In off-axis mode, the biggest improvement comes with the correction of the millimeter-scale defects, specifically diamond-turning signatures on the off-axis parabolas. With this aim, a complete upgrade of the OAPs in GRAVITY's fiber couplers is scheduled for late 2027, in order to improve by a factor x10 the surface error of the OAPs in the Fiber Coupler. This upgrade will replace all current OAPs with newly manufactured mirrors from TNO, based on the development of a specific manufacturing and post-polishing method. The manufacturing is specifically designed for metallic mirrors, as-built in the original GRAVITY Fiber Coupler, which will allow to have an homogeneous thermal expansion coefficient within the Fiber Coupler structure and preserve the optical alignment during cool-down. The new manufacturing and metrology process was validated on a prototype of the M12 mirror (shown in Figure 7), which was manufactured and characterized in 2025, and delivered at MPE. The prototype has demonstrated a tenfold improvement in surface quality compared to the existing GRAVITY mirrors, which validates the manufacturing process. This prototype phase is now followed by the manufacturing of a full set of OAPs, to equip 5 Fiber Coupler units (4 units on GRAVITY and 1 spare unit), which amounts to 35 new mirrors based on the initial Fiber Coupler design [5].

## 5. CONCLUSION

In this work, we identified several effects in GRAVITY's dual-field astrometry. We could compare laboratory measurements of the fiber coupler with on-sky observations, and demonstrated that diamond-turning marks on the off-axis parabolas represent the primary source of systematics in off-axis mode. In on-axis mode, the optical fringes from the roof prism dominate the systematics. This distinction explains the current instrument performance, where GRAVITY achieves superior astrometric precision in off-axis mode.

Our analysis reveals that these high-order systematics, while averaging out over full-night observations, can primarily impact short-exposure astrometry on timescales of minutes. To address these limitations, we propose a two-pronged mitigation strategy. First, the pupil modulation system implemented in GRAVITY during 2022 effectively averages down fringing-related systematics for both observing modes. Second, a comprehensive upgrade of the fiber-coupler mirrors is scheduled for late 2027, replacing current components with new mirrors. The development of a M12 prototype already provides a tenfold improvement in surface quality and validates the new manufacturing process on metallic mirrors needed here. This development will be followed by an upgrade of the GRAVITY instrument, as part of the GRAVITY+ project [10], [11]. This hardware upgrade is expected to go even beyond the 30-100μas dual-field astrometric accuracy of GRAVITY within the next few years.